\documentclass[prl,twocolumn,superscriptaddress,showpacs,floatfix,longbibliography]{revtex4-1}
\usepackage{mathrsfs}
\usepackage{amssymb, amsbsy, amsmath, latexsym, dsfont, array, layout,
graphicx,mathrsfs,color,ulem,bm}

\newcommand{\one}{\mathds{1}}
\newcommand{\ket}[1]{\left|{#1}\right\rangle}
\newcommand{\bra}[1]{\left\langle{#1}\right|}

\begin{document}

\title{Observation of critical phenomena in parity-time-symmetric quantum dynamics}

\author{Lei Xiao}
	\affiliation{Beijing Computational Science Research Center, Beijing 100084, China}
	\affiliation{Department of Physics, Southeast University, Nanjing 211189, China}
\author{Kunkun Wang}
	\affiliation{Beijing Computational Science Research Center, Beijing 100084, China}
	\affiliation{Department of Physics, Southeast University, Nanjing 211189, China}
\author{Xiang Zhan}
	\affiliation{Beijing Computational Science Research Center, Beijing 100084, China}
	\affiliation{Department of Physics, Southeast University, Nanjing 211189, China}
\author{Zhihao Bian}
	\affiliation{Beijing Computational Science Research Center, Beijing 100084, China}
	\affiliation{Department of Physics, Southeast University, Nanjing 211189, China}
\author{Kohei Kawabata}
	\affiliation{Department of Physics, University of Tokyo, 7-3-1 Hongo, Bunkyo-ku, Tokyo 113-0033, Japan}
\author{Masahito Ueda}
	\affiliation{Department of Physics, University of Tokyo, 7-3-1 Hongo, Bunkyo-ku, Tokyo 113-0033, Japan}
	\affiliation{RIKEN Center for Emergent Matter Science (CEMS), Wako, Saitama 351-0198, Japan}
\author{Wei Yi}
	\affiliation{Key Laboratory of Quantum Information, University of Science and Technology of China, CAS, Hefei 230026, China}
	\affiliation{CAS Center For Excellence in Quantum Information and Quantum Physics}
\author{Peng Xue}\email{gnep.eux@gmail.com}
	\affiliation{Beijing Computational Science Research Center, Beijing 100084, China}
	\affiliation{State Key Laboratory of Precision Spectroscopy, East China Normal University, Shanghai 200062, China}

\begin{abstract}
We experimentally simulate non-unitary quantum dynamics using a single-photon interferometric network and study the information flow between a parity-time ($\mathcal{PT}$)-symmetric non-Hermitian system and its environment. We observe oscillations of quantum-state distinguishability and complete information retrieval in the $\mathcal{PT}$-symmetry-unbroken regime.
We then characterize in detail critical phenomena of the information flow near the exceptional point separating the $\mathcal{PT}$-unbroken and -broken regimes, and demonstrate power-law behavior in key quantities such as the distinguishability and the recurrence time. We also reveal how the critical phenomena are affected by symmetry and initial conditions.
Finally, introducing an ancilla as an environment and probing quantum entanglement between the system and the environment, we confirm that the observed information retrieval is induced by a finite-dimensional entanglement partner in the environment.
Our work constitutes the first experimental characterization of critical phenomena in $\mathcal{PT}$-symmetric non-unitary quantum dynamics.
\end{abstract}


\maketitle

Parity-time ($\mathcal{PT}$)-symmetric non-Hermitian systems feature unconventional properties in synthetic systems ranging from classical optical systems~\cite{G+09,R+10,CG11,ACMG12,L+12,B+14,Weimannnm,EPnature,CRDMD10,MMDU15,LYWM13} and microwave cavities~\cite{BSFF14,Bellecnc,ZJSB16} to quantum gases~\cite{leluo} and single photons~\cite{Xue17,dynamicqw}. In these systems, the spectrum is entirely real in the $\mathcal{PT}$-symmetry-unbroken regime, in contrast to the regime with spontaneously broken $\mathcal{PT}$ symmetry~\cite{BB98,BBJ02,B07}. As a result, the dynamics is drastically different in the $\mathcal{PT}$-symmetry-unbroken and -broken regimes, and dynamical criticality occurs at the boundary between the two regimes~\cite{H12,NPreview}. In previous experiments, such unconventional dynamical properties as well as signatures of the $\mathcal{PT}$-transition point, or the exceptional point, were observed in classical $\mathcal{PT}$-symmetric systems with balanced gain and loss~\cite{EPnature,CRDMD10,MMDU15,LYWM13,ACMG12,XMJH16}. Whereas quantum systems with passive $\mathcal{PT}$ symmetry were realized recently~\cite{Xue17,leluo}, critical phenomena in $\mathcal{PT}$-symmetric quantum dynamics are yet to be experimentally explored. Understanding these critical phenomena in the quantum regime provides an important perspective for the study of open quantum systems and is useful for applications in quantum information.

A paradigmatic example of $\mathcal{PT}$-symmetric non-unitary dynamics in the context of open quantum systems is the reversible-irreversible criticality in the information flow between a system and its environment~\cite{KAU17}. Herein, information lost to the environment can be fully retrieved when the system is in the $\mathcal{PT}$-symmetry-unbroken regime because of the existence of a finite-dimensional entanglement partner in the environment protected by $\mathcal{PT}$ symmetry. In contrast, the information flow is irreversible when the system spontaneously breaks $\mathcal{PT}$ symmetry.
Close to the exceptional point, physical quantities such as distinguishability between time-evolved states and the recurrence time of the distinguishability exhibit power-law behavior.

In this work, we simulate $\mathcal{PT}$-symmetric non-unitary quantum dynamics using a single-photon interferometric network, and
experimentally investigate the critical phenomena in the information flow close to the exceptional point.
To extract critical phenomena from non-unitary dynamics, a faithful characterization of the long-time dynamics is necessary. This poses a serious experimental challenge, because maintaining and probing coherent dynamics in the long-time regime is difficult. We overcome this difficulty by directly implementing non-unitary time-evolution operators at any given time, and simulate the non-unitary quantum dynamics by performing non-unitary gate operations on the initial state. We then perform quantum-state tomography on the time-evolved state, which enables us to confirm the critical power-law scaling in various physical quantities. Since our experimental protocol is general enough to implement a broad class of non-unitary operators, we are able to examine in detail the role of symmetry and initial states on non-unitary quantum dynamics driven by a series of related non-Hermitian Hamiltonians.
Furthermore, introducing ancillary degrees of freedom as the environment, we explicitly demonstrate oscillations in the quantum entanglement between the system and the environment in the unitary dynamics of the combined system-environment quantum system. This demonstrates the existence of a finite-dimensional entanglement partner in the environment, which is responsible for the information retrieval. Our work is the first experiment to characterize critical phenomena in $\mathcal{PT}$-symmetric non-unitary quantum dynamics, and opens up an avenue toward simulating the $\mathcal{PT}$-symmetric dynamics in synthetic quantum systems.

\begin{figure*}
\includegraphics[width=0.9\textwidth]{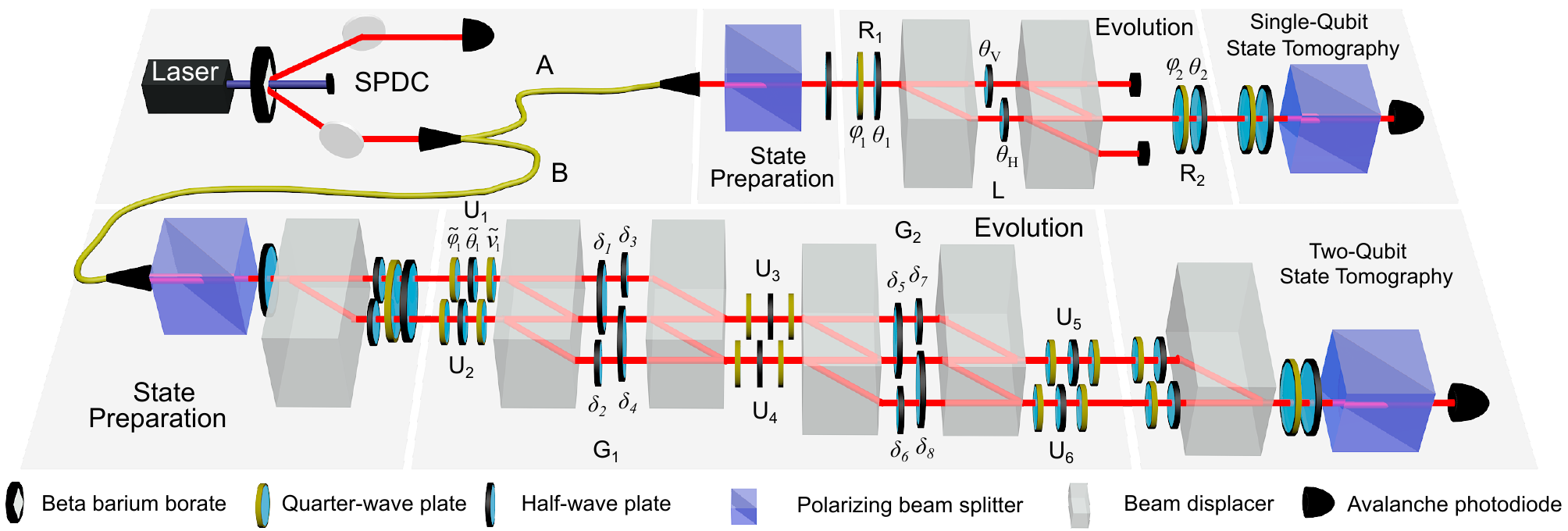}
\caption{Experimental setup. A photon pair is created via spontaneous parametric downconversion (SPDC). One of the photons serves as a trigger, the other is projected into the polarization state $\ket{H}$ or $\ket{V}$ with a polarizing beam splitter (PBS) and a half-wave plat (HWP), and then goes through various optical elements. Upper layer: the experimental setup for the non-unitary dynamics of a single-qubit $\mathcal{PT}$-symmetric system. Lower layer: experimental setup for the unitary dynamics of a two-qubit system, where the two qubits are encoded in polarizational and spatial degrees of freedom.
To prepare the initial state of a two-qubit system, heralded single photons pass through a PBS and a HWP with certain setting
angles and are split by a birefringent calcite beam displacer (BD) into two parallel spatial modes: upper and lower modes. After passing through wave plates inserted into different spatial modes, the photons are prepared in the state $\ket{\bar{0}}$ or $\ket{\bar{1}}$ (see the main text for the definition).
We construct the time-dependent density matrices by performing quantum-state tomography for both of these states.}
\label{fig:setup}
\end{figure*}

{\it Experimental setup:---}
To simulate the dynamics of a two-level $\mathcal{PT}$-symmetric system, we encode basis states in the horizontal and vertical polarizations of a single photon, with $\ket{H}=(1,0)^\text{T}$ and $\ket{V}=(0,1)^\text{T}$.
We generate heralded single photons via type-I spontaneous parametric downconversion, with one photon serving as a trigger and the other as a signal. The signal photon is then projected into the initial state $\ket{H}$ or $\ket{V}$ with a polarizing beam splitter (PBS) and a half-wave plate (HWP), and is sent to the interferometric network as illustrated in Fig.~\ref{fig:setup}.

Experimentally, instead of implementing a non-Hermitian Hamiltonian, we directly realize the time-evolution operator $U$ at any given time $t$ and access time-evolved states by enforcing $U$ on the initial state. As illustrated in Fig.~\ref{fig:setup}, this is achieved by decomposing $U$ according to
\begin{equation}
U=R_2(\theta_2,\varphi_2)L(\theta_H,\theta_V)R_1(\theta_1,\varphi_1),
\label{eqn:Utilde}
\end{equation}
where the rotation operator $R_j(\theta_j,\varphi_j)$ ($j=1,2$) is realized using a quarter-wave plate (QWP) at $\varphi_j$ and a HWP at $\theta_j$, and the polarization-dependent loss operator $L$ is realized by a combination of two beam displacers (BDs) and two HWPs with setting angles $\theta_H$ and $\theta_V$~\cite{supp}.

Here, the setting angles $\{\theta_j,\varphi_j,\theta_H,\theta_V\}$ of wave plates are determined numerically for each given time $t$, such that $U=e^{-i\hat{H}_\text{eff}t}$. In our experiment, the effective non-Hermitian Hamiltonian is given by
\begin{equation}
\hat{H}_\text{eff}=\hat{\sigma}_x+i a(\hat{\sigma}_z-\hat{\one}),
\label{eq:H}
\end{equation}
where $\hat{\sigma}_{x(z)}$ are the standard Pauli operators, and $\hat{\one}$ is the identity operator. The non-Hermitian Hamiltonian $\hat{H}_\text{eff}$ possesses passive $\mathcal{PT}$ symmetry, which can be easily mapped to a $\mathcal{PT}$-symmetric Hamiltonian $\hat{H}_\mathcal{PT}$ with balanced gain and loss, with $\hat{H}_\mathcal{PT}=\hat{H}_\text{eff}+i a\hat{\one}$. Here $a>0$ controls non-Hermiticity, and the Hamiltonian becomes Hermitian for $a=0$; the system is in the $\mathcal{PT}$-symmetry-unbroken (-broken) regime for $0< a<1$ ($a>1$), with the exceptional point located at $a=1$.

The non-unitary dynamics of the system is captured by the time-dependent density matrix~\cite{KAU17, BG12}
\begin{equation}
\rho_{1,2}(t)=\frac{e^{-i\hat{H}_\mathcal{PT}t}\rho_{1,2}(0)e^{i\hat{H}^\dagger_\mathcal{PT}t}}{\text{Tr}\left[e^{-i\hat{H}_\mathcal{PT}t}\rho_{1,2}(0)e^{i\hat{H}^\dagger_\mathcal{PT}t}\right]},
\label{eqn:rhot}
\end{equation}
with the initial density matrices $\rho_{1(2)}(0)=|H(V)\rangle\langle H(V)|$. Note that applying $\hat{H}_\text{eff}$ or $\hat{H}_\mathcal{PT}$ in Eq.~(\ref{eqn:rhot}) would give the same time-dependent matrices. Experimentally, we construct the density matrix at any given time $t$ via quantum-state tomography after signal photons passed through the interferometric setup.
Essentially, we measure the probabilities of photons in the bases $\{\ket{H},\ket{V},\ket{P_+}=(\ket{H}+\ket{V})/\sqrt{2},\ket{P_-}=(\ket{H}-i\ket{V})/\sqrt{2}\}$ through a combination of QWP, HWP, and PBS, and then perform a maximum-likelihood estimation of the density matrix. The outputs are recorded in coincidence with trigger photons. Typical measurements yield a maximum of $18,000$ photon counts over $3$ seconds.

{\it Measuring distinguishability:---}
We characterize information flowing into and out of the system via the trace distance defined by
\begin{align}
D\left[\rho_1(t),\rho_2(t)\right]=\frac{1}{2}\text{Tr}\left|\rho_1(t)-\rho_2(t)\right|,
\end{align}
with $|A|=\sqrt{A^{\dagger}A}$. The trace distance $D$ measures the distinguishability of the two quantum states characterized by $\rho_1(t)$ and $\rho_2(t)$. An increase in the distinguishability signifies information backflow from the environment, whereas a monotonic decrease means unidirectional information flow to the environment~\cite{KAU17, BLP09}.

\begin{figure}
\includegraphics[width=9cm]{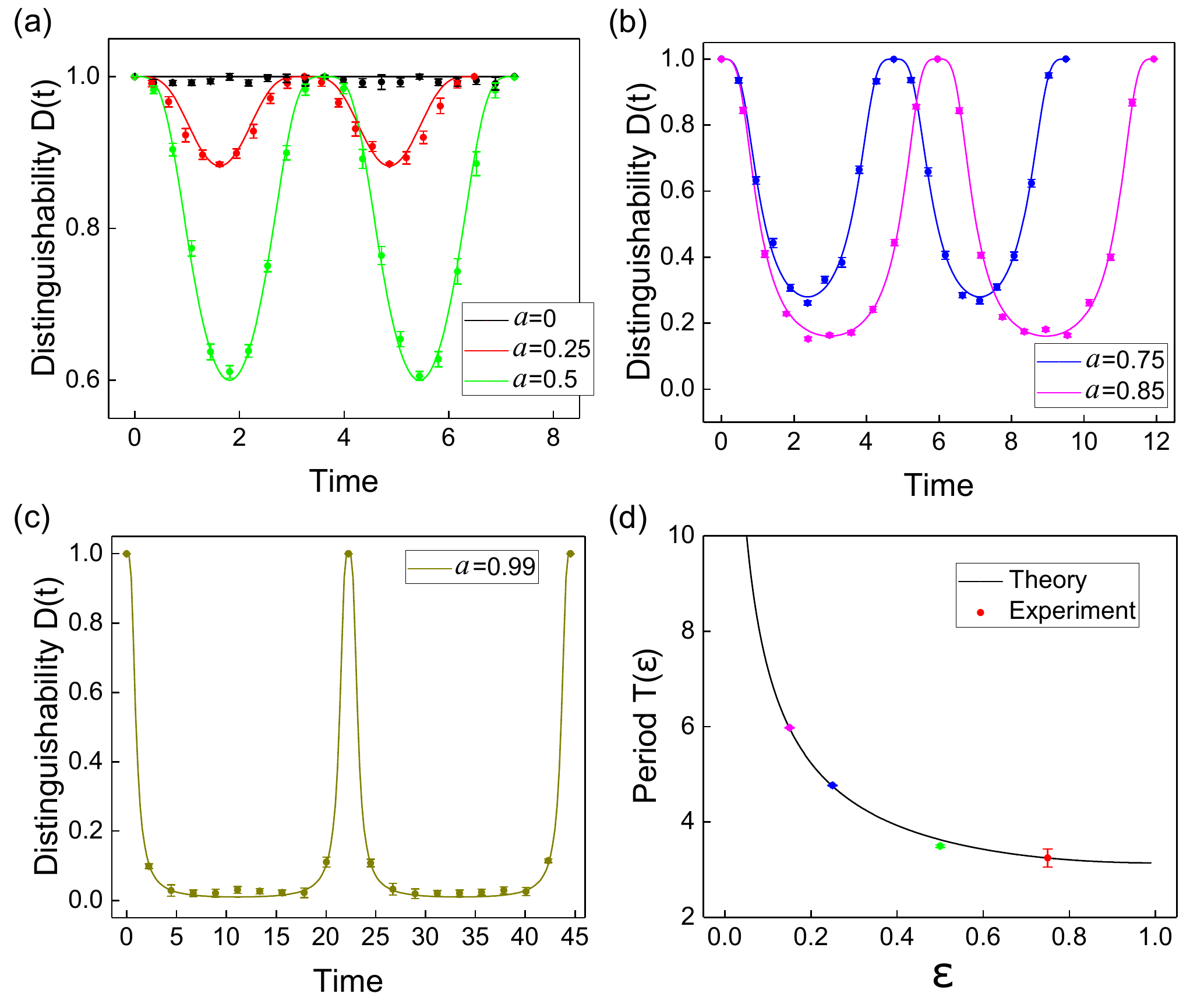}
   \caption{Information retrieval in the $\mathcal{PT}$-symmetry-unbroken regime. (a)-(c) Oscillations of the distinguishability $D(t)$ for $a < 1$, and between the two time-evolved states starting from $\ket{H}$ and $\ket{V}$. Dots with error bars represent the experimental results, while the curves show the theoretical predictions. (d) Recurrence time $T$ of the distinguishability as a function of $\varepsilon=1-a$.
   }
	\label{fig:unbroken}
\end{figure}

In Fig.~\ref{fig:unbroken}, we show the time evolution of the distinguishability when the system is in the $\mathcal{PT}$-symmetry-unbroken regime with $a<1$. For comparison, we also show the case of a unitary evolution with $a=0$. As illustrated in Figs.~\ref{fig:unbroken}(a-c), $D(t)$ oscillates in time when $a<1$, suggesting complete information retrieval with the initial trace distance fully restored periodically. The period of the oscillation $T$, or the recurrence time, increases as the system approaches the exceptional point. We extract the recurrence time by fitting the experimental data with a Fourier series. As shown in Fig.~\ref{fig:unbroken}(d), the recurrence time agrees well with the analytic expression $T=\pi/\sqrt{1-a^2}$~\cite{KAU17}. In the limit $\epsilon\rightarrow 0$ with $\epsilon=1-a$, the recurrence time should diverge as $T\sim \epsilon^{-1/2}$.

\begin{figure}
\includegraphics[width=0.5\textwidth]{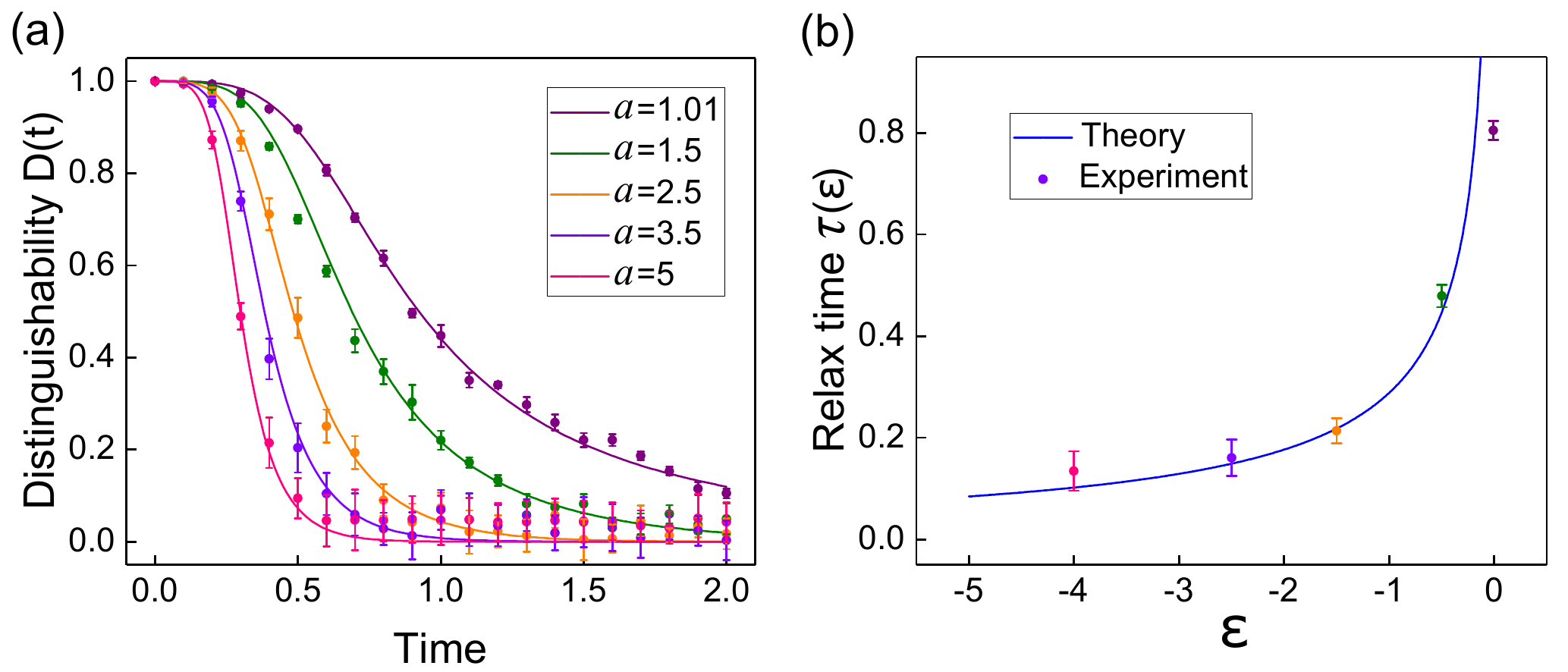}
   \caption{Unidirectional information flow to the environment in the $\mathcal{PT}$-symmetry-broken regime. (a) Decay of the distinguishability in the $\mathcal{PT}$-broken regime with different coefficients $a > 1$. (b) Relaxation time $\tau$ of the distinguishability as a function of $\varepsilon = 1-a$. The blue solid curve shows the theoretical result $\tau=1/2\sqrt{a^2-1}$. We readout the experimental results (dots with error bars) by fitting the experimental data to an exponential function.}
	\label{fig:broken}
\end{figure}

In Fig.~\ref{fig:broken}(a), we show the time evolution of the distinguishability when the system is in the $\mathcal{PT}$-symmetry-broken regime with $a>1$. Here, the distinguishability decays exponentially in time. Fitting the experimental data using $D(t)=D(0)e^{-t/\tau}$, where $D(0)$ is a constant and $\tau$ is the relaxation time, we find that the relaxation time increases as the system approaches the exceptional point. As shown in Fig.~\ref{fig:broken}(b), the measured $\tau$ agrees excellently with the analytical result $\tau=1/2\sqrt{a^2-1}$~\cite{KAU17}, which also diverges with a power-law scaling $\tau \sim |\epsilon|^{-1/2}$ as $\epsilon \rightarrow 0$.

\begin{figure*}
\includegraphics[width=0.75\textwidth]{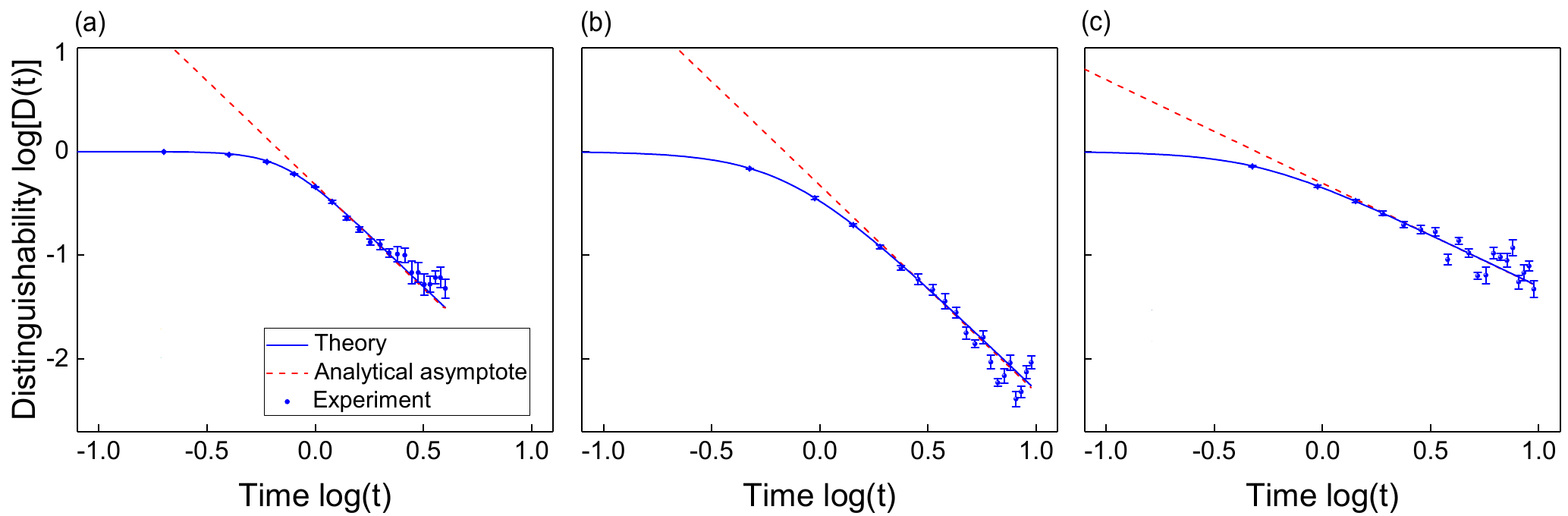}
   \caption{(a) Power-law behavior of the distinguishability $D(t)\sim t^{-2}$  of the $\mathcal{PT}$-symmetric system with initial states $\{\ket{H},\ket{V}\}$. (b)(c) Power-law behavior of the distinguishability of the time-reversal symmetric system at the exceptional point $a=1$ with initial states (b) $(\ket{H}\pm\ket{V})/\sqrt{2}$ or (c) $\{\ket{H},\ket{V}\}$. The power law behaviors  $D(t)\sim t^{-2}$ in (b) and $D(t)\sim t^{-1}$ in (c) demonstrate the dependence of the critical phenomena on the initial state. In the long-time limit, the experimentally measured distinguishability agrees well with theoretical predictions. The blue solid curves show the theoretical prediction, and the red dashed lines indicate their asymptotes.}
	\label{fig:ep}
\end{figure*}

Finally, at the exceptional point ($a=1$), the distinguishability exhibits power-law behavior in the long-time limit. As illustrated in Fig.~\ref{fig:ep}(a), the long-time behavior of the distinguishability agrees well with the theoretical prediction $D(t)\sim t^{-2}$~\cite{KAU17}. Importantly, the observed critical phenomena do not depend on the details of the system but the order of the exceptional point, which signifies their universality~\cite{H12,NPreview}. We note that the measurement suffers from a relatively larger systematic error at long times due to the small $D(t)$.

{\it Symmetry and initial states:---}
Since our experimental protocol is quite general and capable of implementing a broad class of non-unitary operators~\cite{supp}, we are able to investigate the role of symmetry and initial states on the information flow and critical phenomena.
In particular, we experimentally simulate non-unitary dynamics governed by i) $\hat{H}_\mathcal{T}=\hat{\sigma}_{x} + ia \hat{\sigma}_{y}$ and ii) $\hat{H}=\hat{\sigma}_x+(c+ia)\hat{\sigma}_z$. Whereas $\hat{H}_\mathcal{T}$ has time-reversal symmetry $\mathcal{T}\hat{H}_\mathcal{T}\mathcal{T}^{-1}=\hat{H}_\mathcal{T}$ with complex conjugation $\mathcal{T}$, $\hat{H}$ has no relevant symmetries for $a \neq 0$ and $c \neq 0$.

We first study dynamics under $\hat{H}_\mathcal{T}$ with different parameters and initial states $(\ket{H}\pm\ket{V})/\sqrt{2}$.
Since eigenenergies of $\hat{H}_\mathcal{T}$ are given as $\pm\sqrt{1-a^2}$, the exceptional point is located at $a=1$. As shown in Fig.~\ref{fig:ep}(b), the same critical phenomena emerge under time-reversal symmetry: information is retrieved only in the symmetry-unbroken regime $(0<a<1)$~\cite{supp}, and critical scaling is still $D(t)\sim t^{-2}$. However, when we choose the initial states $\{\ket{H},\ket{V}\}$, the critical scaling at the exceptional point is now $D(t)\sim t^{-1}$, as illustrated in Fig.~\ref{fig:ep}(c). This new universality arises because $\ket{H}$ is one of the eigenstates of $\hat{H}_\mathcal{T}$. We note that the same scaling relation can be realized under $\hat{H}_\mathcal{PT}$ with the initial states $(\ket{H}\pm i\ket{V})/\sqrt{2}$ since $(\ket{H}-i\ket{V})/\sqrt{2}$ is one of the eigenstates of $\hat{H}_\mathcal{PT}$.
For the dynamics governed by $\hat{H}$, however, the lack of symmetry therein prevents the information retrieval and the distinguishability decays in time just as in the symmetry-broken cases~\cite{supp}.

{\it $\mathcal{PT}$ dynamics embedded in a two-qubit system:---}
To unveil the origin of the information retrieval, we embed the $\mathcal{PT}$ dynamics into a larger Hilbert space by introducing an ancilla as the environment~\cite{KAU17, GS08}. The combined two-qubit system is governed by the Hermitian Hamiltonian
\begin{equation}
\hat{H}_\text{tot}=\hat{\one}^{\text{a}}\otimes \hat{H}_\text{s}+\hat{\sigma}^{\text{a}}_y\otimes \hat{V}_\text{s},
\label{eq:tot}
\end{equation}
where $\hat{\one}^{\text{a}}$ and $\hat{\sigma}^{\text{a}}_y$ act on the ancilla; $\hat{H}_\text{s}$ and $\hat{V}_\text{s}$ act on the $\mathcal{PT}$-symmetric system with
\begin{align}
&\hat{H}_\text{s}=c^{-1} \left(\hat{H}_\mathcal{PT} \hat{\eta}^{-1}+\hat{\eta}\hat{H}_\mathcal{PT}\right),\\
&\hat{V}_\text{s}=ic^{-1}\left(\hat{H}_\mathcal{PT}-\hat{H}_\mathcal{PT}^\dagger \right).
\end{align}
Here, the non-Hermitian Hamiltonian satisfies pseudo-Hermiticity with
$\hat{\eta}\hat{H}_\mathcal{PT}=\hat{H}_\mathcal{PT}^\dagger\hat{\eta}$ and $\hat{\eta}=\frac{1}{\sqrt{1-a^2}}\begin{pmatrix}
    1 & -ia \\
    ia & 1 \\
    \end{pmatrix}$, where we have $c=\sum_{j}1/\lambda_j$ with the eigenvalues $\lambda_{j}$ of $\hat{\eta}$. Whereas the unitary time evolution of the two-qubit system is driven by the Hermitian Hamiltonian $\hat{H}_\text{tot}$, the effective time evolution of the subsystem is non-unitary and driven by $\hat{H}_\mathcal{PT}$ under a post-selection on the ancilla.

Experimentally, the ancillary degrees of freedom are encoded in two independent spatial modes of the signal photons $\{\ket{u},\ket{d}\}$. Thus, the bases of two-qubit states can be written as
$\{\ket{u}\ket{H}=(1,0,0,0)^\text{T},\ket{u}\ket{V}=(0,1,0,0)^\text{T},\ket{d}\ket{H}=(0,0,1,0)^\text{T},\ket{d}\ket{V}=(0,0,0,1)^\text{T}\}$. The unitary operator $U_\text{tot}$ can be decomposed into~\cite{S77,S82,PW94,IZB+17}
\begin{equation}
U_\text{tot}=\begin{pmatrix}
    U_5 & 0 \\
    0 & U_6 \\
    \end{pmatrix} G_2 \begin{pmatrix}
    U_3 & 0 \\
    0 & U_4 \\
    \end{pmatrix} G_1 \begin{pmatrix}
    U_1 & 0 \\
    0 & U_2 \\
    \end{pmatrix},
\label{eqn:twoqubitmain}
\end{equation}
where $U_j$ ($j=1,\cdots,6$) are unitary single-qubit rotation operators, and $G_1$ and $G_2$ are two-qubit operators. While $U_j$ is implemented using a set of sandwiched wave plates with the configuration QWP (at $\tilde{\phi}_j$ )-HWP (at $\tilde{\theta}_j$)-QWP (at $\tilde{\nu}_j$), $G_{1}$ and $G_{2}$ are realized via a combination of BDs and HWPs with setting angles $\delta_j$ ($j=1,..,8$) as illustrated in Fig.~\ref{fig:setup}~\cite{supp}. Similar to the single-qubit case, these setting angles are determined numerically to ensure that Eq.~(\ref{eqn:twoqubitmain}) yields the correct time evolution operator $U_\text{tot}=e^{-i\hat{H}_\text{tot}t}$.

We then construct the final state at any given time $t$ through quantum-state tomography. This is achieved through measurements of probabilities of photons in $16$ bases given by $\left\{\ket{u},\ket{d},\ket{S_+},\ket{S_-}\right\}\otimes \{\ket{H},\ket{V},\ket{P_+},\ket{P_-}\}$ with $\ket{S_+}=(\ket{u}+\ket{d})/\sqrt{2}$, and $\ket{S_-}=(\ket{u}-i\ket{d})/\sqrt{2}$.
From the measured correlations, we calculate density matrices of the system and the ancilla from
\begin{align}
 \rho^\text{s}_j(t)=\text{Tr}_\text{a}\left[\rho^\text{tot}_j(t)\right],~
 \rho^\text{a}_j(t)=\text{Tr}_\text{s}\left[\rho^\text{tot}_j(t)\right]~(j=1,2).
\end{align}
On the other hand, the density matrix of the $\mathcal{PT}$-symmetric system is
\begin{equation}
\rho^\mathcal{PT}_j(t)=\mathcal{N}\,\text{Tr}_\text{a}\left[\left(\ket{u}\bra{u}\otimes\hat{\one}\right)\rho^\text{tot}_j(t)(\ket{u}\bra{u}\otimes\hat{\one})^\dagger\right],
\end{equation}
with a normalization constant $\mathcal{N}$.

\begin{figure}
\includegraphics[width=0.5\textwidth]{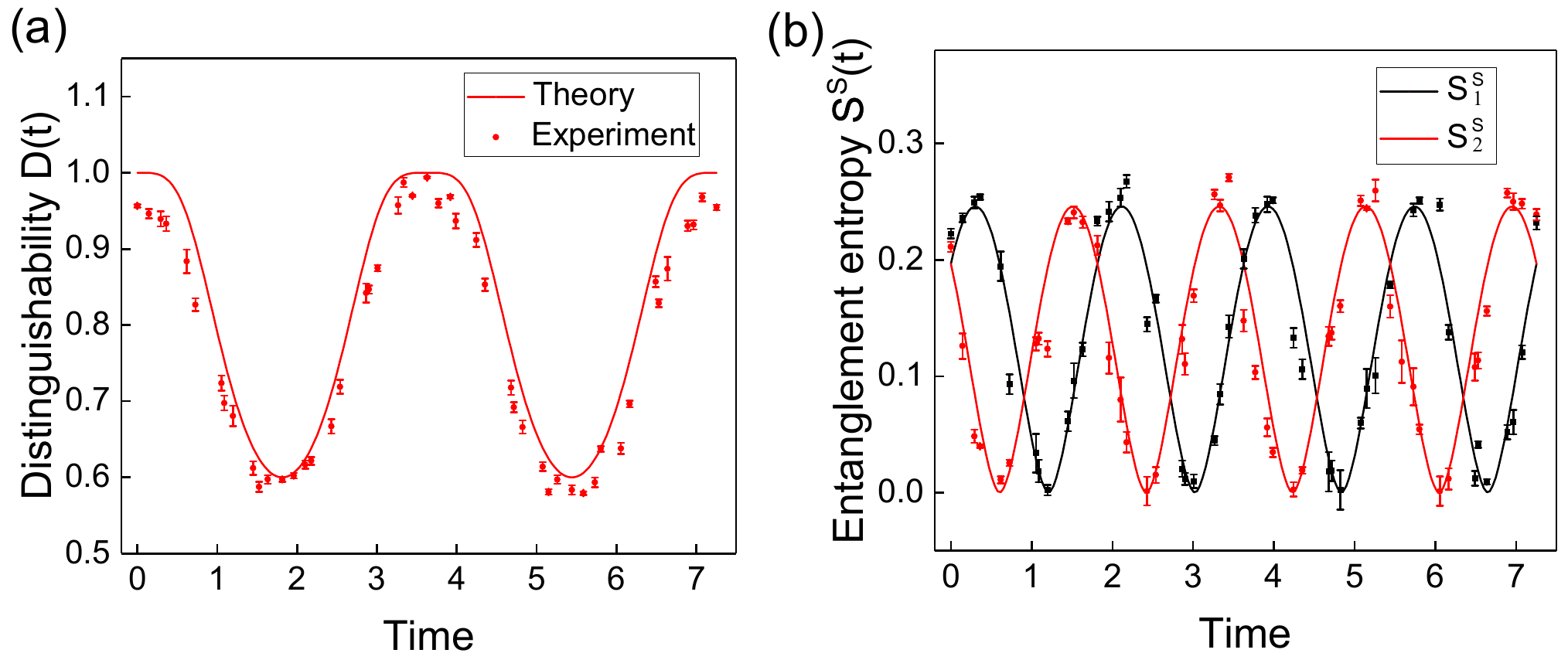}
   \caption{$\mathcal{PT}$ dynamics embedded in a two-qubit system with $a=0.5$. (a)~Quantum oscillations of the distinguishability. (b)~Entanglement entropy between the $\mathcal{PT}$-symmetric system and its ancilla, which oscillates with half the period of that of the distinguishability.}
	\label{fig:twoqubit1}
\end{figure}

For comparison with the single-qubit case, we calculate the distinguishability between two time-evolved states $\rho^\mathcal{PT}_{1}(t)$ and $\rho^\mathcal{PT}_{2}(t)$. We choose the initial states as $\ket{\bar{0}}\propto \ket{u}\ket{H}+\ket{d}\otimes\hat{\eta}\ket{H}$ and $\ket{\bar{1}}\propto\ket{u}\ket{V}+\ket{d}\otimes\hat{\eta}\ket{V}$. As illustrated in Fig.~\ref{fig:twoqubit1}(a), the information retrieval can be observed in the $\mathcal{PT}$-symmetric system.

Importantly, the existence of a hidden entangled partner behind the information retrieval is revealed through the entanglement entropy between the system and the ancilla. The time-dependent entanglement entropy between the $\mathcal{PT}$-symmetric system and its ancilla are calculated as
\begin{equation}
S^\text{s}_j(t)=-\text{Tr}\left[\rho^\text{s}_j(t)\log\rho^\text{s}_j(t)\right].
\end{equation}
In Fig.~\ref{fig:twoqubit1}(b), we show the experimental results of the entanglement entropy for the two different initial states. The entanglement entropy oscillates with half the period of that of the distinguishability, which is consistent with the theoretical predictions~\cite{KAU17} and confirms the exchange of quantum information between the system and the environment during the $\mathcal{PT}$ dynamics.


{\it Conclusion:---}
We have experimentally simulated $\mathcal{PT}$-symmetric quantum dynamics using single-photon interferometric networks. Enforcing non-unitary gate operations on photons and performing quantum-state tomography, we have reconstructed a time-dependent density matrix of the $\mathcal{PT}$ dynamics at arbitrary times. This enables us to characterize critical phenomena close to the $\mathcal{PT}$-transition point and reveal a hidden entanglement partner in the environment. Our work is the first experimental demonstration of critical phenomena in $\mathcal{PT}$-symmetric non-unitary quantum dynamics. We expect that critical phenomena associated with higher-order exceptional points can also be probed using a similar approach.

\begin{acknowledgments} This work has been supported by the Natural Science Foundation of China (Grant Nos. 11674056 and 11522545) and the Natural Science Foundation of Jiangsu Province (Grant No. BK20160024). LX is supported by Postgraduate Research \& Practice Innovation Program of Jiangsu Province. KK is supported by the Japan Society for the Promotion of Science (JSPS) through Program for Leading Graduate Schools (ALPS). MU acknowledges support by KAKENHI Grant Nos. JP18H01145 and JP15H05855 from the JSPS. WY acknowledges support from the National Key Research and Development Program of China (Grant Nos. 2016YFA0301700,2017YFA0304100).
\end{acknowledgments}

\bibliography{reference}

\clearpage
\widetext
\appendix

\renewcommand{\thesection}{\Alph{section}}
\renewcommand{\thefigure}{S\arabic{figure}}
\renewcommand{\thetable}{S\Roman{table}}
\setcounter{figure}{0}
\renewcommand{\theequation}{S\arabic{equation}}
\setcounter{equation}{0}

\section{Supplemental Material for ``Observation of critical phenomena in parity-time-symmetric quantum dynamics''}

\section{Experimental implementation of non-unitary dynamics}

\begin{figure}
\includegraphics[width=0.7\textwidth]{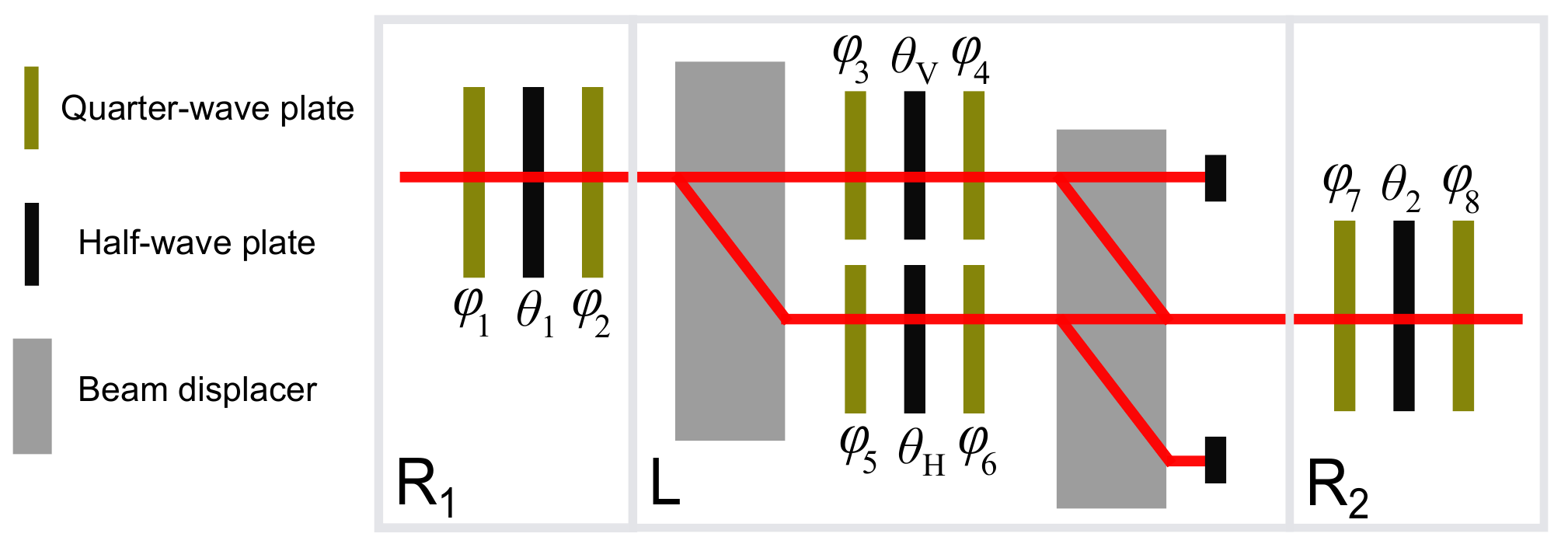}
\caption{Experimental setup to realize a generic class of single-qubit non-unitary operators.}
\label{fig:setupS}
\end{figure}

As illustrated in Fig.~\ref{fig:setupS}, we experimentally implement single-qubit non-unitary operators by realizing
\begin{equation}
U=R_2(\varphi_8,\theta_2,\varphi_7)L(\varphi_6,\theta_H,\varphi_5,\varphi_4,\theta_V,\varphi_3)R_1(\varphi_2,\theta_1,\varphi_1).
\label{eq:u}
\end{equation}
Here, the rotation operators
\begin{align}
&R_1(\varphi_2,\theta_1,\varphi_1)=R_\text{QWP}(\varphi_2)R_\text{HWP}(\theta_1)R_\text{QWP}(\varphi_1), \\ \nonumber
&R_2(\varphi_8,\theta_2,\varphi_7)=R_\text{QWP}(\varphi_8)R_\text{HWP}(\theta_2)R_\text{QWP}(\varphi_7)
\end{align}
are realized by a set of sandwich-type wave plates including two quarter-wave plates (QWPs)
\begin{equation}
R_\text{QWP}(\varphi_j)=\begin{pmatrix}
    \cos\varphi^2_j+i\sin\varphi^2_j & (1-i)\sin\varphi_j\cos\varphi_j \\
    (1-i)\sin\varphi_j\cos\varphi_j & \sin\varphi^2_j+i \cos\varphi^2_j\\
    \end{pmatrix},
\end{equation}
and a half-wave plate (HWP)
\begin{equation}
R_\text{HWP}(\theta_j)=\begin{pmatrix}
    \cos2\theta_j & \sin2\theta_j \\
    \sin2\theta_j & -\cos2\theta_j \\
    \end{pmatrix},
\end{equation}
where $\theta_j$ and $\varphi_j$ are tunable setting angles.

Non-unitarity is introduced through the loss operator $L$ which transmits photons with polarization-dependent transmissivity while flipping their polarizations at the same time. The loss operator is given by
\begin{equation}
L(\varphi_6,\theta_H,\varphi_5,\varphi_4,\theta_V,\varphi_3)=
\begin{pmatrix}
 0 & \xi \\
 \eta & 0 \\
\end{pmatrix},
\end{equation}
where
\begin{align}
&\xi=\frac{1}{2} \left[i \sin2
   \theta_V-\sin 2(\theta _V-\varphi_3)+\sin2(\theta_V-\varphi_4)+i \sin 2(\theta_V-\varphi_3-\varphi_4)\right],\\ 
&\eta=\frac{1}{2} \left[i \sin 2\theta_H+\sin2(\theta_H-\varphi_5)-\sin2(\theta_H-\varphi_6)+i\sin2(\theta_H-\varphi_5-\varphi_6)\right].
\end{align}
In our experiment, we use two beam displacers (BDs) and two sets of wave plates (QWP-HWP-QWP) to implement $L$. The optical axes of BDs are cut so that the vertically polarized photons are transmitted directly and the horizontally polarized photons are displaced into a neighboring spatial mode. Between the two BDs, two sets of wave plates (QWP-HWP-QWP) are inserted into the upper and lower spatial modes and enforce rotations $R$ on the polarization states. Horizontally polarized photons in the upper spatial mode and vertically polarized photons in the lower mode are then combined and transmitted through the second BD, while the other photons are discarded and lost from the system.
This procedure gives the required polarization-dependent photon loss. Note that by tuning the setting angles of the wave plates, we can realize a wide spectrum of non-unitary operators, as detailed below.

To implement a single-qubit non-unitary operator $U=\begin{pmatrix}
A e^{i\alpha} & B e^{i\beta} \\
C e^{i\gamma} & D e^{i\eta} \\
\end{pmatrix}$, we numerically determine the setting angles of the wave plates. For this setup, there are a total of $12$ parameters which can be controlled independently. For an arbitrary $2\times2$ matrix $U$, there are at most $8$ independent parameters. Therefore, in principle, we can use this setup to simulate an arbitrary single-qubit non-unitary operation.

Specifically, for the dynamics of the parity-time ($\mathcal{PT}$)-symmetric system, the time-reversal-symmetric system, and the non-Hermitian system without symmetry, we need to simulate the non-unitary operations with only a few independent parameters. For example, with four QWPs (at $\varphi_3$, $\varphi_4$, $\varphi_5$, and $\varphi_6$) between the BDs, the matrix is simplified as
\begin{equation}
U=R_2(\varphi_8,\theta_2,\varphi_7)L(\theta_H,\theta_V)R_1(\varphi_2,\theta_1,\varphi_1),
\end{equation}
where $L(\theta_H,\theta_V)=
\begin{pmatrix}
 0 & \sin2\theta_V \\
 \sin2\theta_H & 0 \\
\end{pmatrix}$. Furthermore, with the setting angle $\varphi_2=0$ fixed, we obtain $U=\begin{pmatrix}
 Ae^{i\alpha} & Be^{i\beta} \\
 Be^{i\beta} & De^{i\eta} \\
\end{pmatrix}$ with $5$ independent parameters. With $\theta_H=-\theta_V$, the single-qubit non-unitary operator is $U=\begin{pmatrix}
 Ae^{i\alpha} & Be^{i\beta} \\
 Be^{-i\beta} & De^{i\eta} \\
\end{pmatrix}$, where $A$, $B$, $D$, $\alpha$, $\beta$, and $\eta$ are real and independent of each other. With this simplified setup, we can simulate the non-unitary dynamics without symmetry governed by the Hamiltonian $\hat{H}=\hat{\sigma}_x+(c+ia)\hat{\sigma}_z$ in the main text.

If we further remove the two QWPs (at $\varphi_2$ and $\varphi_8$), the matrix in Eq.~(\ref{eq:u}) becomes \begin{equation}
U=R_\text{HWP}(\theta_2)R_\text{QWP}(\varphi_7)L(\theta_H,\theta_V)R_\text{HWP}(\theta_1)R_\text{QWP}(\varphi_1).
\end{equation}
With $\varphi_1=0$, $\theta_1=\theta_2+\pi/4$, and $\varphi_7=2\theta_2$, the single-qubit non-unitary operator is $U=\begin{pmatrix}
 A & iB \\
 iB & D \\
\end{pmatrix}$. With this simplified setup, we can simulate the non-unitary dynamics with both $\mathcal{PT}$-symmetric Hamiltonian $\hat{H}_\mathcal{PT}=\hat{\sigma}_x+i a\hat{\sigma}_z$ and time-reversal-symmetric Hamiltonian $\hat{H}_\mathcal{T}=\hat{\sigma}_{x} + ia \hat{\sigma}_{y}$.

\section{Other symmetries and initial states}

\begin{figure*}
\includegraphics[width=\textwidth]{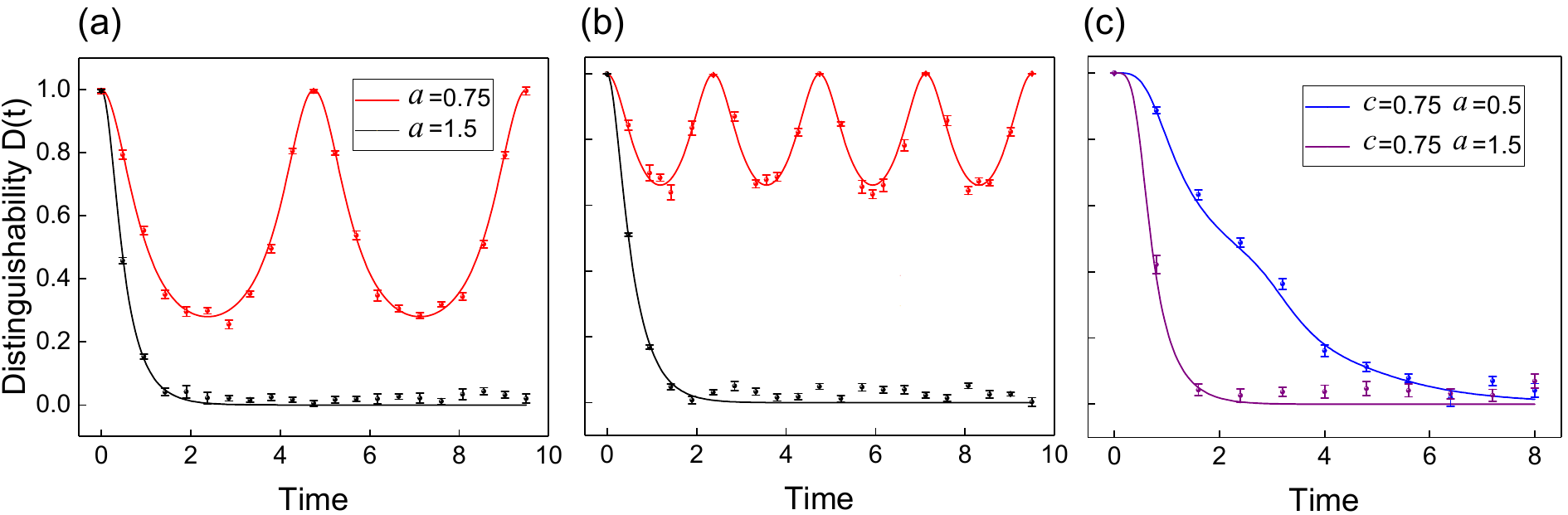}
   \caption{(a) Distinguishability $D(t)$ of the time-reversal-symmetric system governed by $\hat{H}_{\mathcal{T}}$ with initial states $(\ket{H}\pm\ket{V})/\sqrt{2}$. (b) Distinguishability $D(t)$ under $\hat{H}_\mathcal{T}$ with initial states $\{\ket{H},\ket{V}\}$. The distinction between (a) and (b) demonstrates the dependence of the critical phenomena on initial states. (c) Decay of $D(t)$ of the non-Hermitian system governed by $\hat{H}$ (without symmetry) for the two different coefficients with the initial states $\{\ket{H},\ket{V}\}$.}
	\label{fig:new}
\end{figure*}

\begin{figure*}
\includegraphics[width=0.5\textwidth]{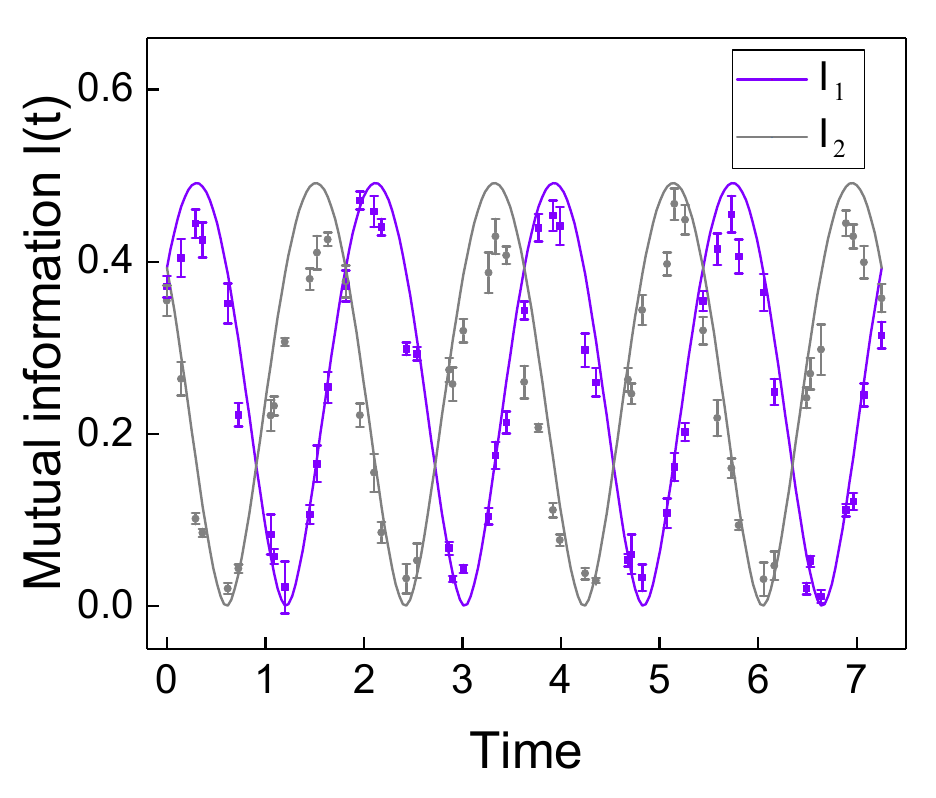}
   \caption{Oscillations of the quantum mutual information $I_j(t)$ between the system and the ancilla with $a=0.5$.}
\label{fig:information}
\end{figure*}

In this section, we present more details on the non-unitary dynamics driven by the Hamiltonians $\hat{H}_\mathcal{T}=\hat{\sigma}_{x} + ia \hat{\sigma}_{y}$ and $\hat{H}=\hat{\sigma}_x+(c+ia)\hat{\sigma}_z$, and with different initial states. Here $\hat{H}_\mathcal{T}$ has time-reversal symmetry and $\hat{H}$ has no relevant symmetries.

As shown in Figs.~\ref{fig:new}(a) and (b), under $\hat{H}_\mathcal{T}$, information is retrieved only in the symmetry-unbroken regime $(0<a<1)$, whereas the period of oscillations in $D(t)$ shows initial-state dependence. The critical scalings at the exceptional point are also different for different initial states, as demonstrated in Figs.~\ref{fig:ep}(b) and (c) of the main text.

For comparison, we show in Fig.~\ref{fig:new}(c) the dynamics under $\hat{H}$, where the lack of symmetry prevents the information retrieval and the distinguishability decays in time just as in the symmetry-broken cases.

%

\section{Experimental implementation of two-qubit evolution}

We embed a $\mathcal{PT}$-symmetric system into a larger Hermitian system $\hat{H}_\text{tot}$ as shown in the main text. Under $\hat{H}_\text{tot}$, the two-qubit state is given by
\begin{equation}
\ket{\Psi_\text{tot}(t)}=\ket{0}\otimes\ket{\psi_\mathcal{PT}(t)}+\ket{1}\otimes\hat{\eta}\ket{\psi_\mathcal{PT}(t)}.
\end{equation}
 Enforcing a post-selected measurement on the ancilla with $\ket{0}$, we obtain the state $\ket{\psi_\mathcal{PT}(t)}$ which is driven by $\hat{H}_\mathcal{PT}$.

As discussed in the main text, we encode the ancilla in the spatial modes of the signal photon. The bases of the two-qubit states are
$\{\ket{u}\ket{H}=(1,0,0,0)^\text{T},\ket{u}\ket{V}=(0,1,0,0)^\text{T},\ket{d}\ket{H}=(0,0,1,0)^\text{T},\ket{d}\ket{V}=(0,0,0,1)^\text{T}\}$, where $\ket{u}$ and $\ket{d}$ denote the upper and lower spatial modes, respectively. The unitary operator $U_\text{tot}$ can be decomposed into~\cite{S77,S82,PW94,IZB+17}
\begin{equation}
U_\text{tot}=\begin{pmatrix}
    U_5 & 0 \\
    0 & U_6 \\
    \end{pmatrix} G_2 \begin{pmatrix}
    U_3 & 0 \\
    0 & U_4 \\
    \end{pmatrix} G_1 \begin{pmatrix}
    U_1 & 0 \\
    0 & U_2 \\
    \end{pmatrix},
\label{eqn:twoqubit}
\end{equation}
where $U_j$ ($j=1,\cdots,6$) is a unitary single-qubit rotation with
\begin{equation}
U_j=R_\text{QWP}(\tilde{\nu}_j)R_\text{HWP}(\tilde{\theta}_j)R_\text{QWP}(\tilde{\phi}_j),
\end{equation}
and $G_1$ and $G_2$ are two-qubit operations with
\begin{align}
G_1&=\begin{pmatrix}
    0 & -\sin2\delta_{31} & 0 & 0\\
    \sin2\delta_{41} & 0 & 0 & \cos2\delta_{41} \\
    \cos2\delta_{41} & 0 & 0 & -\sin2\delta_{41} \\
    0 & 0 & \sin2\delta_{42} & 0 \\
    \end{pmatrix}, \\
G_2&=\begin{pmatrix}
    0 & -\sin2\delta_{75} & 0 & 0\\
    \sin2\delta_{85} & 0 & 0 & \cos2\delta_{85} \\
    \cos2\delta_{85} & 0 & 0 & -\sin2\delta_{85} \\
    0 & 0 & \sin2\delta_{86} & 0 \\
    \end{pmatrix}.
\end{align}
Here, we have $\delta_{jk}=\delta_j-\delta_k$ and $\sin2\delta_{31}=\sin2\delta_{42}=1$. Note that the setting angles $\delta$, $\tilde{\theta}$, $\tilde{\phi}$, and $\tilde{\nu}$ are determined numerically so that the decomposition in Eq.~(\ref{eqn:twoqubit}) holds.

The two-qubit operation
$\begin{pmatrix}
    U_j & 0 \\
    0 & U_{j+1} \\
    \end{pmatrix}$
can be written as $\ket{u}\bra{u}\otimes U_j+\ket{d}\bra{d}\otimes U_{j+1}$ and realized by a single-qubit rotation $U_j$ ($U_{j+1}$) of the polarization states of photons in the upper (lower) mode. Here $G_1$ is given as a single-qubit rotation $\begin{pmatrix}
    \cos2\delta_{41} & -\sin2\delta_{41} \\
    \sin2\delta_{41} & \cos2\delta_{41} \\
    \end{pmatrix}$
on the basis state $\{\ket{u}\ket{H},\ket{d}\ket{V}\}$, and
$\begin{pmatrix}
    0 & -1 \\
    1 &  0 \\
    \end{pmatrix}$
on $\{\ket{u}\ket{V},\ket{d}\ket{H}\}$. These operations can be implemented by two BDs and four HWPs. The BDs split and combine the photons into designated spatial modes depending on their polarizations, and the HWPs implement single-qubit rotations. Moreover, $G_2$ can be implemented in a similar fashion.

\section{Experimental results of the quantum mutual information}

As a further evidence for the exchange of quantum information between the system and the environment, we show in Fig.~\ref{fig:information} experimental results of the quantum mutual information between the system and the ancilla. The quantum mutual information is defined as
\begin{equation}
I_j(t)=S^\text{s}_j(t)+S^\text{a}_j(t)-S^\text{tot}_j(t), \quad (j=1,2)
\end{equation}
with $S^\text{a}_j=-\text{Tr}\left[\rho^\text{a}_j(t)\log\rho^\text{a}_j(t)\right]$ and $S^\text{tot}_j=-\text{Tr}\left[\rho_j^{\text{tot}}(t)\log\rho^\text{tot}_j(t)\right]$. Here $I_1(t)$ ($I_2(t)$) is the quantum mutual information for the quantum dynamics starting from the initial state $\ket{\bar{0}}$ ($\ket{\bar{1}}$).
As shown in Fig.~\ref{fig:information}, the quantum mutual information also oscillates with the period $\pi/(2\sqrt{1-a^2})$, which is the same as that of the entanglement entropy and one-half of that of the quantum-state distinguishability.

\end{document}